\documentclass[fleqn,a4paper]{article}

\usepackage{verbatim,color}

\definecolor{gray}{gray}{0.2}

\newcommand{\CPP}          {\texttt{C$^{++}$}}
\newcommand{\STL}          {\texttt{STL}}

\begin{document}
\raggedright
\sffamily

\title{Multi-dimensional interpolations in \CPP}
\author{M.\ de Jong\\
  {\small NWO-I, Nikhef, PO Box 41882, Amsterdam, 1098 DB Netherlands} \\
  {\small Leiden University, Leiden Institute of Physics, PO Box 9504, Leiden, 2300 RA Netherlands}
}
\maketitle

\begin{abstract}
A \CPP\ software design is presented that can be used to interpolate data in any number of dimensions.
The design is based on a combination of 
templates of functional collections of elements and
so-called type lists.
The design allows for different search methodologies and interpolation techniques in each dimension.
It is also possible 
to expand and reduce the number of dimensions,
to interpolate composite data types and
to produce on-the-fly additional values such as derivatives of the interpolating function.
\end{abstract}

\section{Introduction}

A common problem in numerical computations is the interpolation of data.
Many solutions exist in one dimension, 
several in two dimensions and 
few in any number of dimensions \cite{ref:numerical-recipes}.
An interpolation can be considered as a functional operation, i.e.\ $x \rightarrow y$.
In this, an abscissa value is ``mapped'' onto an ordinate value.
In $n$ dimensions, the functional operation becomes $(x_0, x_1, \ldots, x_{n-1}) \rightarrow y$.
Now, a set of abscissa values is mapped onto an ordinate value.
So, the idea is to use multi-dimensional maps (i.e.\ a map of map of map etc.) 
in which each map corresponds to one dimension in the considered space. 
Such a map should then have the additional functionality required for interpolations.
It will be shown that this functionality can be implemented in \CPP\ 
for a generic template collection of abscissa and ordinate values.
Utilising the concept of type lists, 
this functionality can readily be expanded to any number of dimensions,
without the need to develop additional code.
\newline

In the following, a \CPP\ software design is presented that can be used to interpolate values in any number of dimensions \cite{ref:cpp}.
In this, the code snippets are only indicative.
The names starting with a capital '\verb'J'' refer to the custom implementation and
the sections between two horizontal lines are examples.
In the Examples section, several use-cases are presented.

\section{Collections}

A collection is an object that represents a group of ordered elements. 
The underlying container of a collection is the \verb'std::vector<>' class.
The sequential access to the elements in a collection is thereby provided by 
the corresponding iterators conform the standard template library (\STL).
The basic definition of a collection can be summarised as follows.

\begin{footnotesize}\color{gray}
\begin{verbatim}
  template<class JElement_t,
           class JDistance_t = JDistance<typename JElement_t::abscissa_type> >
  class JCollection :
    public std::vector<JElement_t>
  {
    typedef typename JElement_t::abscissa_type                   abscissa_type;
    typedef typename JElement_t::ordinate_type                   ordinate_type;
    typedef JElement_t                                           value_type;
    typedef JDistance_t                                          distance_type;
  };
\end{verbatim}
\color{black}\end{footnotesize}

In this, 
the template argument \verb'JElement_t' refers to the data type of the elements and
\verb'JDistance_t' to an auxiliary class for the evaluation of the distance between two abscissa values.
The latter is required for interpolations.
To adopt a consistent nomenclature, 
the data type of the elements of a collection should provide for the following type definitions and member functions.

\begin{footnotesize}\color{gray}
\begin{verbatim}
  template<class JAbscissa_t, class JOrdinate_t>
  struct JElement2D {

    typedef JAbscissa_t                 abscissa_type;
    typedef JOrdinate_t                 ordinate_type;

    abscissa_type getX() const;

    const ordinate_type& getY() const;
    ordinate_type& getY();
  };
\end{verbatim}
\color{black}\end{footnotesize}

In this, \verb'X' and \verb'Y' refer to the abscissa and ordinate, respectively.
The data structure \verb'JElement2D' provides for a common template that will be used later and serves here as an example.
In the default distance operator \verb'JDistance', 
the distance is defined by the arithmetic difference between two abscissa values.
This applies to a variety of data types, including all signed primitive data types.

\begin{footnotesize}\color{gray}
\begin{verbatim}
  template<class JAbscissa_t>
  struct JDistance {

    typedef typename JClass<JAbscissa_t>::argument_type    argument_type;

    inline double operator()(argument_type first,
                             argument_type second) const
    {
      return second - first;
    }

    static double precision;
  };

  template<class JAbscissa_t>
  double JDistance<JAbscissa_t>::precision = std::numeric_limits<double>::min();
\end{verbatim}
\color{black}\end{footnotesize}

In this,
the class \verb'JClass' is used to define the proper data type of the arguments  
(i.e.\ a copy for primitive data types and a constant reference otherwise).
The distance evaluates to a positive (negative) value when the second abscissa value 
is larger (smaller) than the first and to zero when they are equal.
For other distances, the user can specify a designated template class. 
The quantity \verb'precision' refers to a minimal distance which effectively evaluates to zero.
It can conveniently be set to a desired value.
For the ordering of the elements in a collection, the comparison of two elements is required.
This represents the equivalence of the less-than operator for the corresponding abscissa values.
The corresponding implementation is provided by the nested class \verb'JComparator'
which uses the specified distance operator.

\begin{footnotesize}\color{gray}
\begin{verbatim}
  class JCollection 
  {
    struct JComparator {

      inline bool operator()(const JElement_t& first, 
                             const JElement_t& second) const
      {
        return this->getDistance(first.getX(), second.getX()) > 0.0;
      }

      inline bool operator()(const JElement_t& element,
                             typename JClass<abscissa_type>::argument_type x) const
      {
        return this->getDistance(element.getX(), x) > 0.0;
      }

      JDistance_t getDistance;
    };

    JDistance_t getDistance;

  protected:
    JComparator compare;
  };
\end{verbatim}
\color{black}\end{footnotesize}

Now, 
the ordering of the elements in the collection and 
the search for the nearest element can consistently be implemented using this comparison operator
and the \STL\ functions \verb'std::sort' and \verb'std::lower_bound', respectively.

\begin{footnotesize}\color{gray}
\begin{verbatim}
  class JCollection 
  {
    void sort()
    {
      std::sort(this->begin(), this->end(), compare);
    }

    const_iterator lower_bound(typename JClass<abscissa_type>::argument_type x) const
    {
      return std::lower_bound(this->begin(), this->end(), x, compare);
    }

    iterator lower_bound(typename JClass<abscissa_type>::argument_type x) 
    {
      return std::lower_bound(this->begin(), this->end(), x, compare);
    }

    ordinate_type& operator[](typename JClass<abscissa_type>::argument_type x)
    {
      iterator i = this->lower_bound(x);

      if (i == this->end() || this->getDistance(x, i->getX()) > distance_type::precision) {
        i = container_type::insert(i, value_type(x, ordinate_type()));
      }

      return i->getY();
    }
  };
\end{verbatim}
\color{black}\end{footnotesize}

The first \verb'lower_bound' member function will be used for interpolations and the second for insertion of elements.
The usual map operator \verb'JCollection::operator[]' provides 
access to the ordinate value at a given abscissa value.
If there is no element in the collection at a zero distance from the given abscissa value, 
a new element is inserted in the collection. 
A collection can thus accordingly be built.
For example:

\hrulefill\\
\begin{footnotesize}\color{gray}
\begin{verbatim}
  JCollection< JElement2D<double, double> > buffer;

  for (double x = -10.0; x < 10.5; x += 1.0) {
    buffer[x] = ..;
  }
\end{verbatim}
\color{black}\end{footnotesize}
\vspace*{-\baselineskip}
\hrulefill\\[\baselineskip]

The determination of the location of the element nearest to a given abscissa value requires $\mathcal{O}(\log{n})$ operations.
For a collection with equidistant elements, this location can directly be computed
based on 
the minimal abscissa value, 
the maximal abscissa value and 
the number of elements in the collection.
To this end, 
the \verb'lower_bound' functions are re-implemented in class \verb'JGridCollection' 
which otherwise simply derives from \verb'JCollection'. 
For convenience, the implementation of the map operator \verb'JCollection::operator[]' is maintained.
\newline

The \verb'JAbstractCollection' constitutes a simple interface to define a sequence of abscissa values for a collection.
The \verb'JSet' and \verb'JGrid' classes provide for an implementation thereof 
and can be used to define a collection with non-equidistant and equidistant abscissa values, respectively.
A variety of configuration functions can now be defined in which the abscissa values are sampled from an abstract collection.

\begin{footnotesize}\color{gray}
\begin{verbatim}
  class JCollection 
  {
    void configure(const JAbstractCollection<abscissa_type>& bounds, ..);
  };
\end{verbatim}
\color{black}\end{footnotesize}

In addition, the \verb'JCollection' class has arithmetic capabilities.
These strictly apply to the ordinate values 
(note that only collections with identical abscissa values can be added or subtracted).
The arithmetic capabilities can be used for interpolations between whole collections (see section Examples).
\newline

The choice of \verb'std::vector' as container rather than \verb'std::map' is motivated by 
the faster I/O (see below) and 
a more efficient implementation of linear coordinate transformations (not discussed here).
When the custom map operator is used, the order of the elements is preserved by design.
If necessary 
(e.g.\ after using one of the modifiers of the \verb'std::vector' class),
the member function \verb'sort' can be used to restore the order.

\section{Maps}

The \verb'JMap' and \verb'JGridMap' classes can be used to extend a collection into more dimensions.

\begin{footnotesize}\color{gray}
\begin{verbatim}
  template<class JKey_t,
           class JValue_t,
           class JDistance_t = JDistance<JKey_t> >
  class JMap :
    public JCollection<JElement2D<JKey_t, JValue_t>, JDistance_t>
  {};

  template<class JKey_t,
           class JValue_t,
           class JDistance_t = JDistance<JKey_t> >
  class JGridMap :
    public JGridCollection<JElement2D<JKey_t, JValue_t>, JDistance_t>
  {};
\end{verbatim}
\color{black}\end{footnotesize}

In this, the data structure \verb'JElement2D' is used as the basic element.
As can be seen from the definition of these classes,
maps are equivalent to collections.
As a result, the implementation of any additional functionality in addition to that of a collection 
can readily be transferred to any number of dimensions.

\section{Multi-dimensional maps}

The \verb'JMultiMap' class constitutes a multi-dimensional map (i.e.\ a map of a map of a map etc.).
In this context, each element corresponds to a point in a multi-dimensional space with an associated function value.
For the definition of a list of maps, the equivalent of a type list is used \cite{ref:modern-cpp-design}.

\begin{footnotesize}\color{gray}
\begin{verbatim}
  template<class JAbscissa_t,
           class JOrdinate_t,
           class JMaplist_t,
           class JDistance_t = JDistance<JAbscissa_t> >
  class JMultiMap;
  {
    enum { NUMBER_OF_DIMENSIONS = JMapLength<JMapList_t>::value };

    void configure(..);
    void insert(..);

    class super_const_iterator;
    class super_iterator;

    super_const_iterator super_begin() const;
    super_const_iterator super_end() const;

    super_iterator super_begin();
    super_iterator super_end();
  };
\end{verbatim}
\color{black}\end{footnotesize}

In this,
the first template argument refers to the abscissa in each dimension and
the second to the overall ordinate.
The template argument 
\verb'JMaplist_t' refers to a list of maps and 
\verb'JDistance' to the distance operator which defines the comparison operator in each dimension.
As follows from the definition, 
the abscissa type and distance operator are common to all dimensions. 
The class \verb'JMapLength' is simply used to determine the number of dimensions.
The map operator \verb'[]' as well as the functions 
\verb'configure' and 
\verb'insert' 
can be used to build a multi-dimensional map.
In analogy with type lists, a list of maps can be defined using the \verb'JMapList' class \cite{ref:modern-cpp-design}.

\begin{footnotesize}\color{gray}
\begin{verbatim}
  template<template<class, class, class> class JHead_t, class JTail_t = JNullType>
  struct JMapList
  {
    typedef JMapList<JHead_t, typename JTail_t::head_list>   head_list;
    typedef typename JTail_t::tail_type                      tail_type;
  };


  template<template<class, class, class> class A = JNullMap,
           template<class, class, class> class B = JNullMap,
           ..
           template<class, class, class> class Z = JNullMap>
  struct JMAPLIST
  {
    typedef JMapList<A, typename JMAPLIST<B,C,..,Z>::maplist>  maplist;
  };
\end{verbatim}
\color{black}\end{footnotesize}

In this, \verb'JNullType' and \verb'JNullMap' are simple placeholders for 
a non-existing data type and non-existing map, respectively.
The type definitions \verb'head_list' and \verb'tail_type' will become handy later.
The data structure \verb'JMAPLIST' provides for a short-hand notation of a list of up to 26 maps.
For example, a 3D-map can be defined and built as follows.

\hrulefill\\
\begin{footnotesize}\color{gray}
\begin{verbatim}
  typedef JMAPLIST<JMap,
                   JMap,
                   JMap>::maplist                JMaplist_t;

  typedef JMultiMap<double, double, JMaplist_t>  JMultimap_t;

  JMultimap_t buffer;

  for (double x = -1.00; x < +1.05; x += 0.10) {
    for (double y = -1.00; y < +1.05; y += 0.10) {
      for (double z = -1.00; z < +1.05; z += 0.10) {
        buffer[x][y][z] = ..;
      }
    }
  }
\end{verbatim}
\color{black}\end{footnotesize}
\vspace*{-\baselineskip}
\hrulefill\\[\baselineskip]

In addition to the \STL\ iterators which are provided by the underlying \verb'std::vector' in each dimension,
a ``super'' iterator is defined which can be used to iterate over all elements in a multi-dimensional map in a single sequence.
A so-called smart pointer (i.e.\ custom implementation of operator \verb'->') is used to represent 
a multi-dimensional key and a single value in the multi-dimensional map.
The abscissa and ordinate values can recursively be accessed using data members \verb'first' and \verb'second', respectively.
For example, the contents of the above 3D-map can be accessed as follows.

\hrulefill\\
\begin{footnotesize}\color{gray}
\begin{verbatim}
  for (JMultimap_t::super_const_iterator i  = buffer.super_begin();
                                         i != buffer.super_end();
                                       ++i) {
    cout << i->first                  << ", "
         << i->second->first          << ", "
         << i->second->second->first  << " -> "
         << i->second->second->second << endl;
  }
\end{verbatim}
\color{black}\end{footnotesize}
\vspace*{-\baselineskip}
\hrulefill\\[\baselineskip]

A multi-dimensional map can thus be viewed as a one-dimensional array of elements with a multi-dimensional key.
In this, the keys are constant and cannot be modified by either ``super'' iterator. 
To be precise, the \verb'const' in the iterator name refers to the ordinate value which is referenced.
In a similar way, the de-reference operator \verb'*' can be applied to a super iterator.
The result represents a multi-dimensional pair (i.e.\ a pair of a pair of a pair etc.) corresponding to a single element in the multi-dimensional map.
For convenience, either super iterator has the member function \verb'getValue' which returns the overall ordinate,
regardless of the number of dimensions.

\section{Functional collections}

The functionality (i.e.\ the ability to evaluate a set of abscissa values to an ordinate value) of a collection is defined by
the template interface \verb'JFunctional'.

\begin{footnotesize}\color{gray}
\begin{verbatim}
  template<class JArgument_t, class JResult_t>
  struct JFunctional
  {
    typedef JArgument_t                                   argument_type;
    typedef JResult_t                                     result_type;

    virtual void compile() = 0;

    virtual result_type evaluate(const argument_type* pX) const = 0;

    static result_type getValue(const JFunctional&   function,
                                const argument_type* pX)
    {
      return function.evaluate(pX);
    }

    static   typename JClass<result_type>::argument_type 
    getValue(typename JClass<result_type>::argument_type value,
	     const argument_type* pX)
    {
      return value;
    }
  };
\end{verbatim}
\color{black}\end{footnotesize}

In this,
\verb'JArgument_t' refers to the argument(s) of the function operator and 
\verb'JResult_t' to the return value.
The function \verb'evaluate' specifies a recursive function call in any dimension.
Any interpolation method should provide for an implementation of this function.
The two static functions \verb'getValue' are helper functions for the interpolation methods 
to properly terminate the recursive function call.
The function \verb'compile' refers to an action which may be required before the evaluation of the function value 
(e.g.\ for the determination of the second derivatives of a cubic-spline interpolation).
\newline

An error may occur during the interpolation.
A solution could be to return a predefined value but if such a solution is not valid, an exception should be thrown. 
Therefore, a simple interface for the handling of these cases is defined.
For this, the nested interface \verb'JExceptionHandler' is defined.
By default, the member function \verb'action' throws the specified exception.

\begin{footnotesize}\color{gray}
\begin{verbatim}
  struct JFunctional
  {
    class JExceptionHandler
    {
      virtual result_type action(const JException& error) const
      {
        throw error;
      }
    };

    void setExceptionHandler(const JSupervisor& supervisor);
  };
\end{verbatim}
\color{black}\end{footnotesize}

In this, 
\verb'JException' derives from \verb'std::exception' and adds storage for a text message.
A custom exception handler can be installed using the function \verb'setExceptionHandler'.
In this, the argument \verb'JSupervisor' refers to a simple place holder for 
a common implementation of the same \verb'JExceptionHandler' interface. 
The \verb'JDefaultResult' exception handler provides for an implementation of this interface 
which does not throw the exception but returns a predefined value instead.

\begin{footnotesize}\color{gray}
\begin{verbatim}
  struct JFunctional
  {
    struct JDefaultResult :
      public JExceptionHandler
    {
      JDefaultResult(const result_type value);

      virtual result_type action(const JException& exception) const
      {
        return defaultResult;
      }

    private:
      result_type defaultResult;
    };
  };
\end{verbatim}
\color{black}\end{footnotesize}

Finally, the \verb'JFunction' interface defines the base class for all functional maps.

\begin{footnotesize}\color{gray}
\begin{verbatim}
  template<class JArgument_t, class JResult_t>
  class JFunction : 
    public virtual JFunctional<JArgument_t, JResult_t>
  {};
\end{verbatim}
\color{black}\end{footnotesize}

In this, the \verb'JFunctional' interface is declared virtual to avoid multiple instances thereof.

\section{Interpolations}

The interpolation methods are defined as template classes 
which derive from the \verb'JCollection' or \verb'JGridCollection' class and
implement the \verb'JFunctional' interface.
In practice, 
the implementation of the virtual function \verb'evaluate' returns a value 
which is obtained by an interpolation between neighbouring elements in 
the collection which are nearest to a given abscissa value.
For this, the function \verb'lower_bound' of the \verb'JCollection' or \verb'JGridCollection' class is used.
For each interpolation method, 
a corresponding map is defined which can be used to extend the interpolation method to other dimensions.
By construction, the \verb'argument_type' of the function \verb'evaluate' corresponds to the \verb'abscissa_type' of the collection.
Usually, but not necessarily, the \verb'result_type' of the function is the same as the \verb'ordinate_type' of the collection.
\newline

In the following, a list of interpolation methods is presented which includes polynomial, cubic-spline and Hermite-spline interpolation.
By design, this list can be expanded.
It is interesting to note that {\em any} data type can be interpolated provided it has arithmetic capabilities.
This makes it possible to simultaneously interpolate a multitude of values and
to determine derivatives or partial integrals of the interpolating function on the fly.

\subsection{Result types}

Usually, the result type of an interpolation is the same as the ordinate type of the collection.
In some cases, more information is desired.
For this purpose, several designated data types for the return value of the function object operator are defined.
The list of possible data types includes 
\verb'JResultHesse',
\verb'JResultPDF', and
\verb'JResultPolynome'.
For these data types, template specialisations exist of the classes used for interpolations.

\begin{footnotesize}\color{gray}
\begin{verbatim}
  template<class JResult_t>
  struct JResultHesse
  {
    JResult_t f;      //!< function value
    JResult_t fp;     //!< first  derivative
    JResult_t fpp;    //!< second derivative
  };

  template<class JResult_t>
  struct JResultPDF
  {
    JResult_t f;      //!< function value
    JResult_t fp;     //!< first derivative
    JResult_t v;      //!< integral <xmin,x]
    JResult_t V;      //!< integral <xmin,xmax>
  };

  template<unsigned int N, class JResult_t>
  struct JResultPolynome
  {
    JResult_t y[N+1]; //!< function and derivative values
  };
\end{verbatim}
\color{black}\end{footnotesize}

The result type 
\verb'JResultHesse' can be used to obtain the function value as well as its first and second derivative and
\verb'JResultPDF' to obtain the function value, its derivative and (partial) integrals.
These result types are available for both polynomial and spline interpolations.
The \verb'JResultPolynome' can be used to obtain the function value and a specified number of derivatives, 
up to the degree of the polynomial function used for interpolation.
An efficient implementation for the underlying algorithm is presented in reference \cite{ref:mdejong-neville}.
It is only available for polynomial interpolation.
All these data structure have arithmetic capabilities.
Hence, their values can be interpolated in any number of dimensions.
Note that these data structures are templates themselves.
So, the additional information could be obtained in any of the dimensions.
For example, it is possible to determine the full Hessian matrix at a given point in space with a single interpolation (see section Examples).

\subsection{Polynomial interpolation}

The template class \verb'JPolintFunction' can be used for polynomial interpolation.
The underlying interpolation code can be obtained from e.g.\ reference \cite{ref:numerical-recipes}.

\begin{footnotesize}\color{gray}
\begin{verbatim}
  template<unsigned int N,
           class JElement_t,
           template<class, class> class JCollection_t,
           class JResult_t,
           class JDistance_t>
  class JPolintFunction :
    public JCollection_t<JElement_t, JDistance_t>,
    public JFunction<typename JElement_t::abscissa_type, JResult_t>
  {
    virtual result_type evaluate(const argument_type* pX) const
    {
      //-> function evaluation
    }

    virtual void compile() 
    {}
  };
\end{verbatim}
\color{black}\end{footnotesize}

The first template parameter \verb'N' 
corresponds to the degree of the polynomial function which is used to interpolate the data.
The function \verb'compile' does nothing as there is no action needed before the interpolation.
Template specialisations exist for the return types presented above.
There are also template specialisations for \verb'N = 0' and \verb'1'.
The first specialisation acts as a simple look-up table and the second provides for a linear interpolation.
These template specialisations are defined as follows.

\begin{footnotesize}\color{gray}
\begin{verbatim}
  template<class JElement_t,
           template<class, class> class JCollection_t,
           class JResult_t,
           class JDistance_t>
  class JPolintFunction<(0|1), 
                        JElement_t,
                        JCollection_t,
                        typename JResultType<typename JElement_t::ordinate_type>::result_type,
                        JDistance_t> :
    public JCollection_t<JElement_t, JDistance_t>,
    public JFunction<typename JElement_t::abscissa_type,
                     typename JResultType<typename JElement_t::ordinate_type>::result_type>
  {
    public:

    typedef JCollection_t<JElement_t, JDistance_t>                     collection_type;

    typedef typename collection_type::abscissa_type                    abscissa_type;
    typedef typename collection_type::ordinate_type                    ordinate_type;

    typedef typename collection_type::const_iterator                   const_iterator;
    typedef typename collection_type::iterator                         iterator;
    typedef typename collection_type::distance_type                    distance_type;

    typedef typename JResultType<ordinate_type>::result_type           data_type;
    typedef JFunction<abscissa_type, data_type>                        function_type;

    typedef typename function_type::argument_type                      argument_type;
    typedef typename function_type::result_type                        result_type;
  };
\end{verbatim}
\color{black}\end{footnotesize}

In this, 
\verb'(0|1)' refers to the degree of the polynomial function which is used to interpolate the data.
The implementation of the function \verb'evaluate' of the linear interpolation is presented here to serve as an example.

\begin{footnotesize}\color{gray}
\begin{verbatim}
  class JPolintFunction<1, 
                        JElement_t,
                        JCollection_t,
                        typename JResultType<typename JElement_t::ordinate_type>::result_type,
                        JDistance_t>
  {
    virtual result_type evaluate(const argument_type* pX) const
    {
      if (this->size() <= 1u) {
        return this->getExceptionHandler().action(JFunctionalException("..."));
      }

      const argument_type x = *pX;

      const_iterator p = this->lower_bound(x);

      if ((p == this->begin() && this->getDistance(x, (p++)->getX()) > distance_type::precision) ||
          (p == this->end()   && this->getDistance((--p)->getX(), x) > distance_type::precision)) {

        return this->getExceptionHandler().action(JValueOutOfRange("..."));
      }

      ++pX;  // next argument value


      const_iterator q = p--;

      const double dx = this->getDistance(p->getX(), q->getX());
      const double a  = this->getDistance(x, q->getX()) / dx;
      const double b  = 1.0 - a;

      ya = function_type::getValue(p->getY(), pX);
      yb = function_type::getValue(q->getY(), pX);

      ya *= a;
      yb *= b;

      ya += yb;

      return ya;
    }

  private:
    mutable result_type ya;
    mutable result_type yb;
  };
\end{verbatim}
\color{black}\end{footnotesize}

In this, 
\verb'JFunctionalException' and
\verb'JValueOutOfRange' correspond to the error condition in which
there are too few data for the interpolation and 
an abscissa value that is out of the range of the collection, respectively 
(\verb'"..."' refers to an appropriate text message).
The class \verb'JResultType' is an auxiliary class which is used to consistently define 
the result type based on the ordinate type of the collection 
which could constitute an interpolation method by itself.
The function \verb'getValue' is used to properly terminate the recursive function call. 
The use of mutable data members \verb'ya' and \verb'yb' reduces the number of 
temporary copies which improves performance for large composite values.  
\newline

In case the \verb'JResultPDF' is specified as return type, 
the data type of the element of the collection should provide for the storage capacity of the partial integrals.
To this end, the data structure \verb'JPolintElement2S' is defined which extends \verb'JElement2D'.

\subsection{Spline interpolation}

The template class \verb'JSplineFunction' can be used for cubic-spline interpolation.
The underlying interpolation code can be obtained from e.g.\ reference \cite{ref:numerical-recipes}.

\begin{footnotesize}\color{gray}
\begin{verbatim}
  template<class JElement_t,
           template<class, class> class JCollection_t,
           class JResult_t,
           class JDistance_t>
  class JSplineFunction :
    public JCollection_t<JElement_t, JDistance_t>,
    public JFunction<typename JElement_t::abscissa_type, JResult_t>
  {
     virtual result_type evaluate(const argument_type* pX) const
     {
       //-> function evaluation
     }

     virtual void compile()
     {
       //-> calculation of second derivatives
     }
  };
\end{verbatim}
\color{black}\end{footnotesize}

The function \verb'compile' is used to determine the second derivatives for the spline interpolation.
The actual data type of the element of the collection should provide for the storage capacity of these values.
In case the \verb'JResultPDF' is specified as return type, 
the data type should also provide for the storage capacity of the partial integrals.
To this end, the data structures 
\verb'JSplineElement2D' and
\verb'JSplineElement2S' 
are defined, 
which can be used for the cubic-spline interpolation without and with partial integrals, respectively.

\subsection{Hermite-spline interpolation}

The template class \verb'JHermiteSplineFunction' can be used for spline interpolation 
of a monotonously increasing or decreasing data set.

\begin{footnotesize}\color{gray}
\begin{verbatim}
  template<class JElement_t,
           template<class, class> class JCollection_t,
           class JResult_t,
           class JDistance_t>
  class JHermiteSplineFunction :
    public JCollection_t<JElement_t, JDistance_t>,
    public JFunction<typename JElement_t::abscissa_type, JResult_t>
  {
     virtual result_type evaluate(const argument_type* pX) const
     {
       //-> function evaluation
     }

     virtual void compile()
     {
       //-> calculation of first derivatives
     }
  };
\end{verbatim}
\color{black}\end{footnotesize}

The function \verb'compile' is used to determine the first derivatives for the spline interpolation.
The actual data type of the element of the collection should provide for the storage capacity of these values.
In case the \verb'JResultPDF' is specified as return type, 
the data type should also provide for the storage capacity of the partial integrals.
The data structures 
\verb'JSplineElement2D' and
\verb'JSplineElement2S' 
can be used for the Hermite-spline interpolation without and with partial integrals, respectively.

\section{Interpolations in one dimension}

For interpolations in one dimension, the \verb'JFunction1D' interface provides for 
a common implementation of the function object operator \verb'()'.

\begin{footnotesize}\color{gray}
\begin{verbatim}
  template<class JArgument_t, class JResult_t>
  struct JFunction1D :
    public JFunction<JArgument_t, JResult_t>
  {
    enum { NUMBER_OF_DIMENSIONS = 1 };

    result_type operator()(const argument_type x) const
    {
      return this->evaluate(&x);
    }
  };
\end{verbatim}
\color{black}\end{footnotesize}

The interpolation classes in one dimension are simply defined as follows.

\begin{footnotesize}\color{gray}
\begin{verbatim}
  template<unsigned int N,
           class JElement_t,
           template<class, class> class JCollection_t,
           class JResult_t,
           class JDistance_t = JDistance<typename JElement_t::abscissa_type> >
  class JPolintFunction1D :
    public JPolintFunction<N, JElement_t, JCollection_t, JResult_t, JDistance_t>,
    public JFunction1D<typename JElement_t::abscissa_type, JResult_t> 
  {};

  template<class JElement_t,
           template<class, class> class JCollection_t,
           class JResult_t,
           class JDistance_t = JDistance<typename JElement_t::abscissa_type> >
  class JSplineFunction1D :
    public JSplineFunction<JElement_t, JCollection_t, JResult_t, JDistance_t>,
    public JFunction1D<typename JElement_t::abscissa_type, JResult_t> 
  {};

  template<class JElement_t,
           template<class, class> class JCollection_t,
           class JResult_t,
           class JDistance_t = JDistance<typename JElement_t::abscissa_type> >
  class JHermiteSplineFunction1D :
    public JHermiteSplineFunction<JElement_t, JCollection_t, JResult_t, JDistance_t>,
    public JFunction1D<typename JElement_t::abscissa_type, JResult_t> 
  {};
\end{verbatim}
\color{black}\end{footnotesize}

Note that the second and first derivatives of the data 
that are required in member function \verb'evaluate' of classes \verb'JSplineFunction' and \verb'JHermiteSplineFunction', respectively 
are determined beforehand in function \verb'compile'.

\section{Constant function}

A special case of a 1D-function is provided by the \verb'JConstantFunction1D' class.
The return value is a constant which is set at construction.

\begin{footnotesize}\color{gray}
\begin{verbatim}
  template<class JArgument_t, class JResult_t>
  class JConstantFunction1D : 
    public JFunction1D<JArgument_t, JResult_t>
  {
    JConstantFunction1D(const result_type y);

    virtual result_type evaluate(const argument_type* pX) const
    {
      return y;
    }

  private:
    result_type y;
  };
\end{verbatim}
\color{black}\end{footnotesize}

By utilising the arithmetic capabilities of collections,
a constant function can be used to lower the number of dimensions of a multi-dimensional function object (see section Examples).
For this, the type definitions of \verb'head_list' and \verb'tail_type' of the \verb'JMapList' class are used.

\section{Functional maps}

The functional maps are simple definitions which follow from functional collections.

\begin{footnotesize}\color{gray}
\begin{verbatim}
  template<unsigned int N,
           class JKey_t,
           class JValue_t,
           template<class, class, class> class JMap_t,
           class JResult_t,
           class JDistance_t = JDistance<JKey_t> >
  class JPolintMap :
    public JPolintFunction<N, 
                           JElement2D<JKey_t, JValue_t>, 
                           JMapCollection<JMap_t>::template collection_type, 
                           JResult_t,
                           JDistance_t>
  {};

  template<class JKey_t,
           class JValue_t,
           template<class, class, class> class JMap_t,
           class JResult_t,
           class JDistance_t = JDistance<JKey_t> >
  class JSplineMap :
    public JMap_t<JKey_t, JValue_t, JDistance_t>,
    public JFunction<JKey_t, JResult_t>
  {
    virtual result_type evaluate(const argument_type* pX) const
    {
       //-> function evaluation using internal one-dimensional cubic-spline interpolator
    }

    virtual void compile()
    {
       //-> pre-allocate memory for internal one-dimensional cubic-spline interpolator
    }
  };

  template<class JKey_t,
           class JValue_t,
           template<class, class, class> class JMap_t,
           class JResult_t,
           class JDistance_t = JDistance<JKey_t> >
  class JHermiteSplineMap :
    public JMap_t<JKey_t, JValue_t, JDistance_t>,
    public JFunction<JKey_t, JResult_t>
  {
    virtual result_type evaluate(const argument_type* pX) const
    {
       //-> function evaluation using internal one-dimensional Hermite-spline interpolator
    }

    virtual void compile()
    {
       //-> pre-allocate memory for internal one-dimensional Hermite-spline interpolator
    }
  };
\end{verbatim}
\color{black}\end{footnotesize}

In this, the template parameter
\verb'JKey_t' refers to the key of the map,
\verb'JValue_' to the mapped value and
\verb'JMap_t' to the type of map (\verb'JMap' or \verb'JGridMap').
The auxiliary class \verb'JMapCollection' is used here to simply define the appropriate collection type for a given map type.
The second and first derivatives of the data 
that are used in functions \verb'JSplineFunction' and \verb'JHermiteSplineFunction', respectively  
should now be determined during interpolation.
To this end, the function \verb'evaluate' is re-implemented in classes \verb'JSplineMap' and \verb'JHermiteSplineMap'.
A corresponding one-dimensional interpolation method is then applied to the interpolated data from the lower dimension(s).
\newline

To make the functional maps syntactically consistent with the above definition of a multi-dimensional map,
simple wrapper classes are defined, namely:

\begin{footnotesize}\color{gray}
\begin{verbatim}
  template<class JKey_t, class JValue_t, class JDistance_t = JDistance<JKey_t> >
  struct JSplineFunctional[Grid]Map :
    public JSplineMap<JKey_t,
                      JValue_t,
                      J[Grid]Map,
                      typename JResultType<JValue_t>::result_type,
                      JDistance_t>
  {};

  template<class JKey_t, class JValue_t, class JDistance_t = JDistance<JKey_t> >
  struct JPolint(0|1|2|3)Functional[Grid]Map : 
    public JPolintMap<(0|1|2|3),
                      JKey_t,
                      JValue_t,
                      J[Grid]Map,
                      typename JResultType<JValue_t>::result_type,
                      JDistance_t>
  {};

  template<class JKey_t, class JValue_t, class JDistance_t = JDistance<JKey_t> >
  struct JSplineFunctional[Grid]MapH :
    public JSplineMap<JKey_t,
                      JValue_t,
                      J[Grid]Map,
                      JResultHesse<typename JResultType<JValue_t>::result_type>,
                      JDistance_t>
  {};

  template<class JKey_t, class JValue_t, class JDistance_t = JDistance<JKey_t> >
  struct JPolint(0|1|2|3)Functional[Grid]MapH : 
    public JPolintMap<(0|1|2|3),
                      JKey_t,
                      JValue_t,
                      J[Grid]Map,
                      JResultHesse<typename JResultType<JValue_t>::result_type>,
                      JDistance_t>
  {};
\end{verbatim}
\color{black}\end{footnotesize}
 
In this, 
\verb'[Grid]' refers to the optional use of a \verb'JGridMap' instead of \verb'JMap' and 
\verb'(0|1|2|3)' to the degree of the polynomial function which is used to interpolate the data.
The letter '\verb'H'' refers to a functional map which also provides the first and second derivatives of 
the interpolated value in the corresponding dimension. 
\newline

With the availability of 
a variety of interpolation methods,
different types of collections and
template functional maps,
various multi-dimensional interpolation methods can be constructed
with different search methodologies and interpolation techniques in each dimension.

\section{Multi-dimensional interpolations}

The multi-dimensional interpolation is based on the template \verb'JMultiFunction' class.
It provides for an implementation of the function object operator \verb'()'.

\begin{footnotesize}\color{gray}
\begin{verbatim}
  template<class JFunction_t,
           class JMaplist_t,
           class JDistance_t = JDistance<typename JFunction_t::argument_type> >
  class JMultiFunction :
    public JMultiMap<typename JFunction_t::argument_type,
                     JFunction_t,
                     JMaplist_t,
                     JDistance_t>
  {
    enum { NUMBER_OF_DIMENSIONS = JMapLength<JMaplist_t>::value
                                + JFunction_t::NUMBER_OF_DIMENSIONS };

    result_type operator()(..) const;

    void compile();
  };
\end{verbatim}
\color{black}\end{footnotesize}

The way to read the template definition is that of 
a function ``\verb'JFunction_t''' as 
a function of ``\verb'JMaplist_t'''.
The overall number of dimensions \verb'NUMBER_OF_DIMENSIONS' is simply the sum of the two. 
The abscissa and return types follow from the function definition.
Note that the return type can be expanded by any of the functional maps in the given list.
As can be seen from this definition, the  \verb'JMultiFunction' class derives from a  \verb'JMultiMap'.
Hence, the building and the I/O of a multi-dimensional function is the same as that of a multi-dimensional map.
This is convenient, in particular when the acquisition of data requires a significant amount of CPU time.
For instance, a multi-dimensional map can be stored on disk and 
different interpolation methods can subsequently be applied to the same data 
to purposely optimise the balance between speed and accuracy.
The function operator simply diverts the arguments to the recursive interpolation procedure   
which is implemented in the first functional map in the given map list and then
recursively transferred to the other (read ``lower'') dimensions until it terminates in the specified function.
The number of arguments should therefore match the total number of dimensions.
The implementation of the function operator could either be based on 
a variadic function template (since \verb'c++11') or
the ellipsis operator.
The function \verb'compile' applies to all dimensions.
By definition, the axes corresponding to the different dimensions are orthogonal.
The maintenance of abscissa values in each map allows for different axes in a given dimension,  
which may also depend on the abscissa values in the upper dimensions.
This also applies for equidistant abscissa values.
For example, the complete solid angle can effectively be described by different but equidistant 
sets of zenith and azimuth angles.
\newline

It is interesting to note that the template argument \verb'JFunction_t' could also refer to 
an interpolation method in more than one dimension.
In that case, the number of dimensions in which the interpolation works is literally the sum
of the number of dimensions of the specified function and the number of maps in the specified list.
The  
building, 
I/O and
function call 
are then the same as that of a multi-dimensional function defined for the total number of dimensions (see also section Examples).

\section{I/O}

The I/O of multi-dimensional maps can be realised from 
the availability of the I/O of the basic element and the template collection.
To this end, the following member functions for the template collection are defined.

\begin{footnotesize}\color{gray}
\begin{verbatim}
  class JCollection 
  {
    friend inline JReader& operator>>(JReader& in, JCollection& collection)
    {
      int n;

      in >> n;

      this->resize(n);

      for (iterator i = this->begin(); i != this->end(); ++i) {
        in >> *i;
      }

      return in;
    }

    friend inline JWriter& operator<<(JWriter& out, const JCollection& collection)
    {
      const int n = this->size();

      out << n;

      for (const_iterator i = this->begin(); i != this->end(); ++i) {
        out << *i;
      }

      return out;
    }
  };
\end{verbatim}
\color{black}\end{footnotesize}

In this,
\verb'JReader' and
\verb'JWriter' refer to some auxiliary classes which provide for the input and output streaming of binary data, respectively.
In the input stream operation,
the required memory for the given number of elements is allocated at once and 
the order of the elements is maintained.
This significantly speeds up the reading of a large number of elements and collections thereof.
\newline

The I/O of the basic element is also provided.

\begin{footnotesize}\color{gray}
\begin{verbatim}
  struct JElement2D {
    friend inline JReader& operator>>(JReader& in, JElement2D& element)
    {
      in >> element.x;
      in >> element.y;

      return in;
    }

    friend inline JWriter& operator<<(JWriter& out, const JElement2D& element)
    {
      out << element.x;
      out << element.y;

      return out;
    }
  };
\end{verbatim}
\color{black}\end{footnotesize}

Because 
multi-dimensional functions are derived from 
multi-dimensional maps which are based on 
template collections of elements,
the I/O is thereby provided for all function objects. 
Note that the additional data in the extended elements (e.g.\ \verb'JPolintElement2S', \verb'JSplineElement2D' and \verb'JSplineElement2S') 
are not subject to I/O.
Instead, these values are computed in function \verb'compile'. 
Furthermore, 
the I/O of a \verb'JGridCollection' is compatible with that of a \verb'JCollection'.
As a result, 
the I/O of multi-dimensional functions is independent of the search methodology and interpolation technique 
in the different dimensions.

\section{Examples}

\subsection{5D interpolation}

In the following, an example for a 5D-polynomial interpolation method is presented.

\hrulefill\\
\begin{footnotesize}\color{gray}
\begin{verbatim}
  inline double f5(const double x0,
                   const double x1,
                   const double x2,
                   const double x3,
                   const double x4);

  template<class JKey_t, class JValue_t, class JDistance_t>
  struct JMap_t :
    public JPolintMap<3, JKey_t, JValue_t, JGridMap,
                      typename JResultType<JValue_t>::result_type, JDistance_t>
  {};

  typedef JPolintFunction1D<3, JElement2D<double, double>, JGridCollection>   JFunction1D_t;

  typedef JMAPLIST<JMap_t,
                   JMap_t,
                   JMap_t,
                   JMap_t>::maplist                                           JMaplist_t;

  JMultiFunction<JFunction1D_t, JMaplist_t> g5;
  
  // initialisation
 
  for (double x0 = xmin; x0 <= xmax + 0.5*dx; x0 += dx) {
    for (double x1 = xmin; x1 <= xmax + 0.5*dx; x1 += dx) {
      for (double x2 = xmin; x2 <= xmax + 0.5*dx; x2 += dx) {
        for (double x3 = xmin; x3 <= xmax + 0.5*dx; x3 += dx) {
          for (double x4 = xmin; x4 <= xmax + 0.5*dx; x4 += dx) {
            g5[x0][x1][x2][x3][x4] = f5(x0, x1, x2, x3, x4);
          }
        }
      }
    }
  }
    
  g5.compile();
  g5.setExceptionHandler(new JFunction1D_t::JDefaultResult(0.0));

  // evaluation

  const double v = f5(x0,x1,x2,x3,x4);
  const double w = g5(x0,x1,x2,x3,x4);
\end{verbatim}
\color{black}\end{footnotesize}
\vspace*{-\baselineskip}
\hrulefill\\[\baselineskip]

In this, \verb'f5' corresponds to some arbitrary test function.
Here, 
a 1D-interpolation method \verb'JFunction1D_t' is expanded to 
a 5D-interpolation method using 
a 4D-list of interpolating maps \verb'JMaplist_t'.
The data structure \verb'JMap_t' constitutes a mere type definition.
The object \verb'g5' can be built using the map operator, 
where each abscissa value \verb'x(0|1|2|3|4)' enclosed between brackets \verb'[]' corresponds to a specific dimension. 
The object \verb'g5' can be used to evaluate the function value with the same syntax as the test function \verb'f5'.
The values of \verb'v' and \verb'w' are thereby comparable. 
In this example, an interpolation is made using a third degree polynomial in each dimension.
It is interesting to note that the I/O of the interpolator of this example  
is compatible with that of the following two examples.
In other words, an interpolator can be built in one way, stored to disk and subsequently used in many other ways.

\subsection{2D $+$ 3D $=$ 5D interpolation}

In the following, an example for a 2D-polynomial interpolation method is presented 
which is expanded to a 5D-polynomial interpolation method.

\hrulefill\\
\begin{footnotesize}\color{gray}
\begin{verbatim}
  typedef JMAPLIST<JMap_t>::maplist                                           JMaplist1D_t;
  typedef JMAPLIST<JMap_t,
                   JMap_t,
                   JMap_t>::maplist                                           JMaplist3D_t;
  
  typedef JMultiFunction<JFunction1D_t, JMaplist1D_t>                         JFunction2D_t;
  typedef JMultiFunction<JFunction2D_t, JMaplist3D_t>                         JFunction5D_t;

  JFunction5D_t g5;
\end{verbatim}
\color{black}\end{footnotesize}
\vspace*{-\baselineskip}
\hrulefill\\[\baselineskip]

Here, the 1D-interpolation method of the first example \verb'JFunction1D_t' is replaced by 
a 2D-interpolation method \verb'JFunction2D_t' which is subsequently expanded to 
a 5D-interpolation method using 
a 3D-list of interpolating maps.
The same functional maps \verb'JMap_t' are used as before.
The syntax for the initialisation and evaluation of the interpolator is the same as before 
and so are the results.
As can be seen from this example, 
one can transparently add dimensions to existing interpolators.

\subsection{5D $-$ 3D $=$ 2D interpolation}

In the following, an example for a 5D-polynomial interpolation method is presented 
which is decomposed into a 3D-polynomial and a 2D-polynomial interpolation method.

\hrulefill\\
\begin{footnotesize}\color{gray}
\begin{verbatim}
  typedef JMAPLIST<JMap_t>::maplist                                           JMaplist1D_t;
  typedef JMultiFunction<JFunction1D_t, JMaplist1D_t>                         JFunction2D_t;

  typedef JFunction1D_t::abscissa_type                                        abscissa_type;
  typedef JFunction1D_t::value_type                                           value_type;
  typedef JMap<abscissa_type, JCollection<value_type> >                       JMap2D_t;
  typedef JConstantFunction1D<abscissa_type, JMap2D_t>                        JConstant2D_t;

  typedef JMAPLIST<JMap_t,
                   JMap_t,
                   JMap_t,
                   JMap_t>::maplist                                           JMaplist4D_t;
  typedef JMAPLIST<JMap_t,
                   JMap_t,
                   JMap_t>::maplist                                           JMaplist3D_t;

  JMultiFunction<JConstant2D_t, JMaplist3D_t>  g5;

  // evaluation

  JFunction2D_t g2;

  copy(g5(x0,x1,x2), g2);      // put interpolated data into 2D functional object

  g2.compile();

  const double v = f5(x0,x1,x2,x3,x4);
  const double w = g2(x3,x4);
\end{verbatim}
\color{black}\end{footnotesize}
\vspace*{-\baselineskip}
\hrulefill\\[\baselineskip]

Here, the 5D-interpolation method of the first example is reduced to 
a 2D-interpolation method \verb'JFunction2D_t' using 
a 3D-interpolation method returning a 2D-table of interpolated function values.
It should be emphasized that the whole 2D-table is interpolated at once in calling \verb'g5(x0,x1,x2)'.
The syntax for the initialisation of the interpolator is the same as before.
The syntax for the evaluation is now different 
because the return value of the 3D-interpolator is a 2D-table.
Note that if the return value of the 3D-interpolator would be a 2D-interpolator, 
it would act as a 5D-interpolator (see second example).
Hence, the use of function \verb'copy'.
As can be seen from this example, 
one can effectively subtract dimensions from existing interpolators.
This can significantly speed up the process of a large number of interpolations 
in which the abscissa values in some dimensions are unchanged.

\subsection{Hessian matrix}

In the following, an example for a 3D-polynomial interpolation method is presented 
which can be used to determine the Hessian matrix at a given point in space.
The degree of the polynomial is commonly set to \verb'N = 3'. 

\hrulefill\\
\begin{footnotesize}\color{gray}
\begin{verbatim}
  const int N = 3;

  template<class JKey_t, class JValue_t, class JDistance_t>
  struct JMap_t :
    public JPolintMap<N, JKey_t, JValue_t, JGridMap,
                      JResultHesse<typename JResultType<JValue_t>::result_type>,
		      JDistance_t>
  {};

  typedef JPolintFunction1D<N,
                            JElement2D<double, double>,
                            JGridCollection,
                            JResultHesse<double> >          JFunction1D_t;
  typedef JMAPLIST<JMap_t,
                   JMap_t>::maplist                         JMaplist_t;

  typedef JMultiFunction<JFunction1D_t, JMaplist_t>         JMultiFunction_t;
  typedef JMultiFunction_t::result_type                     result_type;
    
  JMultiFunction_t g3;

  for (double x = xmin; x < xmax + 0.5*dx; x += dx) {
    for (double y = xmin; y < xmax + 0.5*dx; y += dx) {
      for (double z = xmin; z < xmax + 0.5*dx; z += dx) {
        g3[x][y][z] = f3(x,y,z);
      }
    }
  }

  const result_type result = g3(x,y,z);
\end{verbatim}
\color{black}\end{footnotesize}
\vspace*{-\baselineskip}
\hrulefill\\[\baselineskip]

In this, \verb'f3' corresponds to some arbitrary test function.
The object \verb'g3' can be built as usual using the map operator, 
where the abscissa values \verb'x', \verb'y' and \verb'z' correspond to the three dimensions.
Because the result type \verb'JResultHesse' (see above) is specified for 
the 1D-interpolation method \verb'JPolintFunction1D' and 
each interpolating map \verb'JPolintMap', 
the function value as well as the first and second derivatives thereof will be determined in each dimension.
The overall result type is then recursively expanded during this process. 
The function value as well as the components of the Hessian matrix at a given position \verb'(x,y,z)' can thus be obtained as follows.

\begin{center}
\renewcommand{\baselinestretch}{2.5}
\begin{small}
  \begin{tabular}{c@{}c@{}c@{}}

    \begin{math}\begin{array}[t]{ccl}
        \\
        \hline
        \frac{\partial^2 f}{\partial x^2}          & \equiv &  \verb'result.fpp.f.f'  \\
    \end{array}\end{math}                                                                    &

    \begin{math}\begin{array}[t]{ccl}
        f                                          & \equiv &  \verb'result.f.f.f'    \\
        \hline
        \frac{\partial^2 f}{\partial x\partial y}  & \equiv &  \verb'result.fp.fp.f'  \\
        \frac{\partial^2 f}{\partial y^2}          & \equiv &  \verb'result.f.fpp.f'    
    \end{array}\end{math}                                                                    &
    
    \begin{math}\begin{array}[t]{ccl|}
        \\
        \hline
        \frac{\partial^2 f}{\partial x\partial z}  & \equiv &  \verb'result.fp.f.fp'  \\
        \frac{\partial^2 f}{\partial y\partial z}  & \equiv &  \verb'result.f.fp.fp'  \\
        \frac{\partial^2 f}{\partial z^2}          & \equiv &  \verb'result.f.f.fpp'  \\
    \end{array}\end{math}                                                                    \\
  \end{tabular}
\end{small}
\end{center}

\section{Conclusions and outlook}

A \CPP\ software design is presented that can be used to interpolate data in any number of dimensions.
It is based on a combination of type lists and templates of one-dimensional functional collections of elements.
The design allows for different search methodologies and interpolation techniques in each dimension.
It is also possible 
to expand and reduce the number of dimensions,
to interpolate composite data types and
to produce on-the-fly additional values such as derivatives of the interpolating function.
It is interesting to note that the design can readily be extended to other functionalities such as filling of histograms and evaluation of integrals, 
in any number of dimensions, that is. 
Because the functional collections are type defined,
the evaluation of integrals can one-to-one be coupled to a suitable method.
For example, 
the degree of a Gauss-Legendre quadrature can be matched to 
the degree of the polynomial interpolation.

\clearpage
\bibliographystyle{plain}
\bibliography{Interpol}

\end{document}